\documentclass[aps,twocolumn,prl,amsmath,amssymb,10pt,aps]{revtex4-1} 

\usepackage[colorlinks=true,linkcolor=blue,citecolor=blue,urlcolor=blue]{hyperref}
\usepackage{graphicx}
\usepackage{latexsym}
\usepackage{amsmath}
\usepackage{wasysym}
\usepackage{amsthm}
\usepackage{amsbsy}
\usepackage{amssymb}
\usepackage{epstopdf} 
\usepackage{enumerate}
\usepackage{setspace} 
\usepackage{dcolumn} 
\usepackage{bm} 
\usepackage{slashed}
\usepackage{color}
\usepackage{youngtab}
\usepackage{float}
\usepackage{graphicx}
\usepackage{multirow}
\usepackage{array}
\usepackage{mathtools}
\usepackage{inputenc}
\usepackage{color}
\usepackage{makecell}

\begin{document}

\title{Adiabatic Path from Fractional Chern Insulators to the Tao-Thouless State}

\author{Sutirtha Mukherjee}
\author{Kwon Park}
\affiliation{School of Physics, Korea Institute for Advanced Study, Seoul 02455, Korea}

\date{\today}

\begin{abstract}
In view of the evolution from the integer to fractional quantum Hall effect, the next frontier in the research of topological insulators is to investigate what happens in fractionally filled topological flat bands. 
A particularly pressing question is if there exists the lattice analogue of the Laughlin state in the 1/3-filled Chern flat band, dubbed as the Chern-Laughlin state. 
The answer depends crucially on the form of the electron-electron interaction, which can generate various competing ground states such as the Laughlin, stripe/nematic, parafermion, and parton states.
Unfortunately, it is difficult to precisely characterize the exact ground state as any of these candidate ground states due to the lack of appropriate order parameters.
Here, we propose that the existence of an adiabatic path from fractional Chern insulators to the Tao-Thouless state, i.e., the root partition state of the Laughlin state in the thin torus limit, can serve as an effective order parameter for the Chern-Laughlin state. 
Specifically, by devising the piecewise hybrid adiabatic path of first transforming the electron-electron interaction and then taking the thin torus limit, it is shown that Chern flat bands with the nearest-neighbor interaction can indeed host the Chern-Laughlin state at 1/3 filling.
This method can be extended to possible FCIs at other general fillings of the Jain sequence.        
\end{abstract}

\maketitle

An ordered phase of matter such as ferromagnet is usually characterized by the order parameter related with symmetry. 
Chern (or generally topological) insulators belong to a different class of the ordered phase, which is characterized by the topological invariant called the Chern number. 
Chern insulators are the lattice analogue of the integer quantum Hall state (IQHS) occurring in fully filled Landau levels (LLs). 

The fractional quantum Hall state (FQHS)~\cite{Tsui82} is a strongly-correlated topological phase of matter occurring in fractionally filled LLs, which cannot be characterized by either symmetry-related order parameter or topological invariant. 
This creates a problem for the characterization of fractional Chern insulators (FCIs), which are envisioned as the lattice analogue of the FQHS~\cite{Sheng11,Neupert11,Regnault11, Wu12,Wu12_Gauge-fixed,Wu12_Adiabatic}.

Immediately after the discovery of the Laughlin state~\cite{Laughlin83}, there were several theoretical attempts~\cite{Girvin87,Read89} to devise the order parameter characterizing the FQHS by drawing an analogy from the Bose-Einstein condensate.
Later, the FQHS has become better understood via the composite fermion (CF) theory~\cite{Jain89}, where CFs form the IQHS in effective CF LLs.
Unfortunately, we do not yet know how to actually compute the Chern number of such effective CF LLs.

An alternative is to compute the so-called many-body Chern number~\cite{Sheng11,Neupert11}, which can take fractional values if the ground state is topologically degenerate.
Considering that the Laughlin state forms the ground state manifold (GSM) with a triple topological degeneracy, the similar GSM in the 1/3-filled Chern flat band
is typically taken as the evidence for the lattice analogue of the Laughlin state, dubbed as the Chern-Laughlin state.

While definitely important, however, the topological degeneracy alone cannot really distinguish between various competing ground states such as the Laughlin, stripe/nematic~\cite{Koulakov96_PRL,Koulakov96_PRB,Xia11}, parafermion~\cite{Jeong17}, and parton~\cite{Faugno21} states.
Is the 1/3-filling FCI truly the Chern-Laughlin state, not any of the other competing states?

Here, we propose that the existence of an adiabatic path from FCIs to the Tao-Thouless state~\cite{Tao83} can serve as an effective order parameter for the Chern-Laughlin state.
The Tao-Thouless state is the root partition state of the Laughlin state in the thin torus limit~\cite{Rezayi94,Bergholtz08}.
The key idea is that, if truly described by the Chern-Laughlin state, the 1/3-filling FCI should evolve adiabatically to the Tao-Thouless state in the thin torus limit.

Unfortunately, the GSM collapses if the thin torus limit is taken directly in Chern flat bands with the nearest-neighbor interaction.
Here, we devise a piecewise hybrid path by first transforming the electron-electron interaction from the nearest-neighbor to Coulomb interaction and then taking the thin torus limit, during which the GSM is shown to remain intact. 
It is confirmed that the exact Coulomb ground state in the 1/3-filled Chern flat band indeed evolves adiabatically to the Tao-Thouless state. 
The Chern-Laughlin state is constructed by incorporating appropriate quantum fluctuations into the Tao-Thouless state, resulting in high overlaps with both exact Coulomb and nearest-neighbor interaction ground states in the 2D bulk, the latter case of which is consistent with previous results~\cite{Wu12_Gauge-fixed}.

\begin{figure*}
\includegraphics[width=1.8\columnwidth,angle=0]{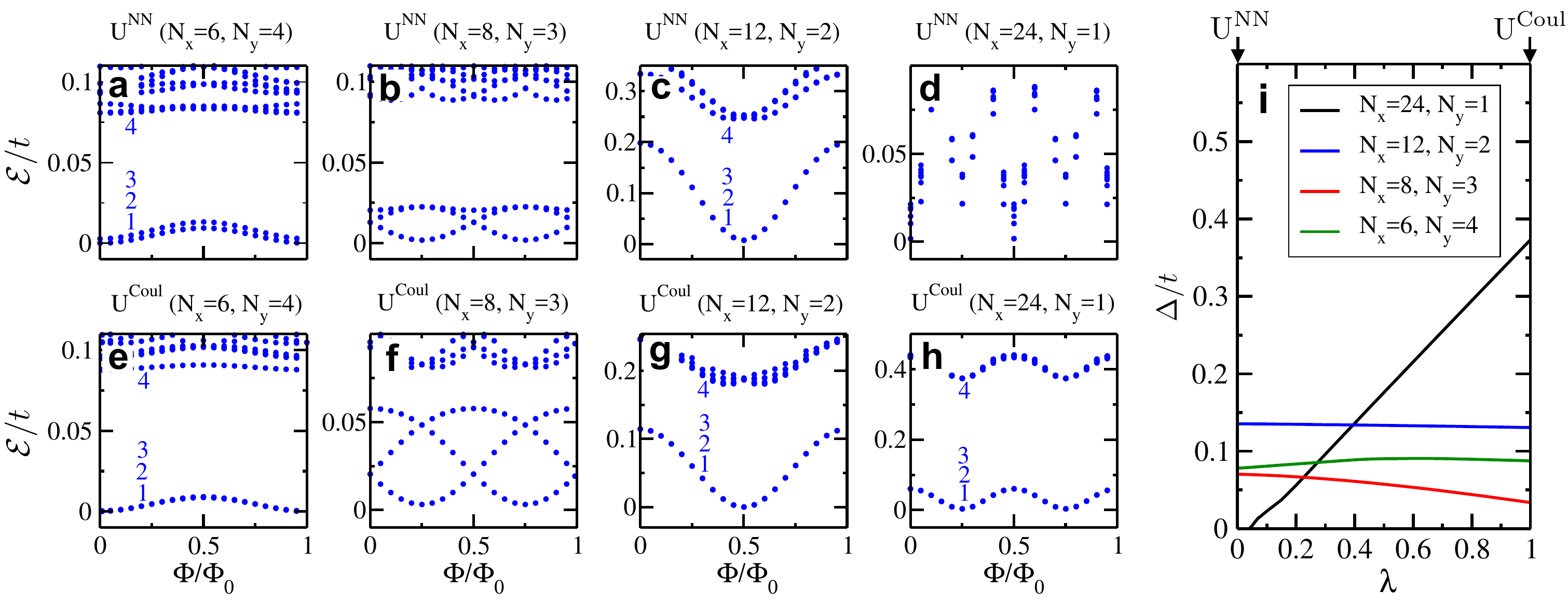}
\caption{
{\bf Comparison between the energy spectra of the nearest-neighbor and Coulomb interactions.}
({\bf a}\mbox{-}{\bf d}) Energy spectra of the nearest-neighbor interaction, $U^{\rm NN}$, in the Chern flat band of the checkerboard lattice model as a function of the test flux, $\Phi/\Phi_0$, for various different values of $(N_x,N_y)$.
$\Phi_0$ is the magnetic flux quantum.
({\bf e}\mbox{-}{\bf h}) Similar energy spectra of the Coulomb interaction, $U^{\rm Coul}$.
Here, the electron and site numbers are $N=8$ and $N_s=N_x N_y=24$, respectively.
Energies are given in units of the nearest-neighbor hopping amplitude $t$ and offset by the lowest energy in each panel.
The first four lowest energies are specified by numbers when three copies of the ground state have almost the same energy. 
({\bf i}) Energy gap $\Delta$ between the third and fourth lowest energies as a function of the mixing parameter $\lambda$, tuning the range of interaction via $U(\lambda)=(1-\lambda)U^{\rm NN}+\lambda U^{\rm Coul}$.
Here, $\Phi$ is set to be zero.
}
\label{fig:spectrum}
\end{figure*}

\noindent{\bf Strongly correlated Chern flat bands}

We begin by considering generic microscopic models for strongly correlated Chern flat bands:
\begin{align}
H= \sum_{\alpha,{\bf k}} \epsilon_{\alpha {\bf k}} c^\dagger_{\alpha {\bf k}} c_{\alpha {\bf k}} +\sum_{i<j} U_{ij} n_i n_j ,
\end{align}
where the energy dispersion $\epsilon_{\alpha {\bf k}}$ is given as a function of the band index $\alpha$ and the Bloch momentum ${\bf k}$.
The electron-electron interaction $U_{ij}$ is initially taken as the nearest-neighbor interaction, but later extended to the Coulomb interaction. 

Depending on system parameters, 
some of the energy bands can be topologically non-trivial and nearly flat, i.e., Chern flat bands.
Here, we focus on two tight-binding models with one in the checkerboard lattice~\cite{Sun11} and the other in the kagome lattice~\cite{Tang11}.
For simplicity, we provide technical details of these models in Supplemental Material.

The final Hamiltonian of interest is the projected Hamiltonian, 
\begin{align}
{\cal H}_{\rm CFB}={\cal P}_{\rm CFB} H {\cal P}_{\rm CFB}, 
\end{align}
where ${\cal P}_{\rm CFB}$ is the projection operator to a Chern flat band. 
After the projection, the energy dispersion is dominated by the electron-electron interaction, whose matrix elements are provided in Supplemental Material.
It is, however, important to note that Chern flat bands are not strictly flat.
Throughout this work, we keep the band dispersion, which turns out to play an important role in the thin torus limit.


We solve ${\cal H}_{\rm CFB}$ at 1/3 filling via exact diagonalization, whose results for the checkerboard lattice are provided in the main text, while those for the kagome lattice are in Supplemental Material.

\noindent{\bf Results}

\noindent{\bf Adiabatic path.}
We first check what happens to energy spectra if one takes the thin torus limit directly in Chern flat bands with the nearest-neighbor interaction, $U^{\rm NN}_{ij}=U \delta_{\langle i,j \rangle}$. 
Throughout this work, we set $U=t$ with $t$ being the nearest-neighbor hopping amplitude in the checkerboard lattice model.
The thin torus limit is achieved by increasing the aspect ratio of the system, $r_a$, while fixing the total number of sites, $N_s$, and that of electrons, $N$.
Specifically, $r_a=N_x/N_y$, while $N_s=N_x N_y=3N$ at 1/3 filling.

Figure~\ref{fig:spectrum}~{\bf a}\mbox{-}{\bf d} show the energy spectra of the nearest-neighbor interaction in the checkerboard lattice model as a function of $(N_x,N_y)$ for $N=8$ and $N_s=24$.
As one can see, the triple-degenerate GSM persists up to $(N_x,N_y)=(12,2)$, but collapses at $(N_x,N_y)=(24,1)$. 
The GSM collapses since the energy gap induced by the nearest-neighbor interaction becomes exceedingly small even when the energy dispersion is completely ignored~\cite{Bernevig12}.
Such a small energy gap can be easily wiped out by nearly, but not completely flat band dispersions.

To establish an adiabatic path from FCIs to the Tao-Thouless state, it is necessary to find a right form of the electron-electron interaction that can sustain the energy gap in the entire process of taking the thin torus limit. 
To this end, we consider the Coulomb interaction, which can be written in the checkerboard lattice as follows:
\begin{align}
U^{\rm Coul}_{ij} = \frac{e^2}{\epsilon} \sum_{l_1, l_2} \frac{1}{| {\bf r}_{ij}+l_1 N_x \hat{x}+l_2 N_y \hat{y} |},
\end{align}
which ${\bf r}_{ij}$ is the relative position vector between the $i$-th and $j$-th sites. 
It is important to note that the Coulomb interaction includes all contributions from the repeated images of electrons induced by the periodic boundary condition~\cite{Yoshioka84}. 
Throughout this work, we set $e^2/\epsilon =U$.

Figure~\ref{fig:spectrum}~{\bf e}\mbox{-}{\bf h} show the energy spectra of the Coulomb interaction in the checkerboard lattice model.
As one can see, the triple-degenerate GSM remains robust all the way to the thin torus limit.  
This shows that the long-range nature of the Coulomb interaction is important to stabilize the adiabatic path to the Tao-Thouless state.

Motivated by this finding, we devise a piecewise hybrid path connecting between FCIs in the 2D bulk and the Tao-Thouless state in the thin torus limit. 
Our path is composed of two pieces; (i) transforming the electron-electron interaction from the nearest-neighbor to Coulomb interaction in the 2D bulk and (ii) taking the thin torus limit with the Coulomb interaction. 

Our path is entirely adiabatic if the first piece of the path is adiabatic since the second is already shown to be so.
To this end, we check if the triple-degenerate GSM remains intact during the transformation process of the electron-electron interaction via $U(\lambda)=(1-\lambda)U^{\rm NN}+\lambda U^{\rm Coul}$ with $\lambda \in [0,1]$. 
Figure~\ref{fig:spectrum}~{\bf i} shows that the energy gap between the third and fourth lowest energies remains open as a function of $\lambda$ in the 2D bulk. 

Next, we would like to confirm that the ground state is actually equal to the Tao-Thouless state in the thin torus limit. 

\begin{figure}
\includegraphics[width=\columnwidth,angle=0]{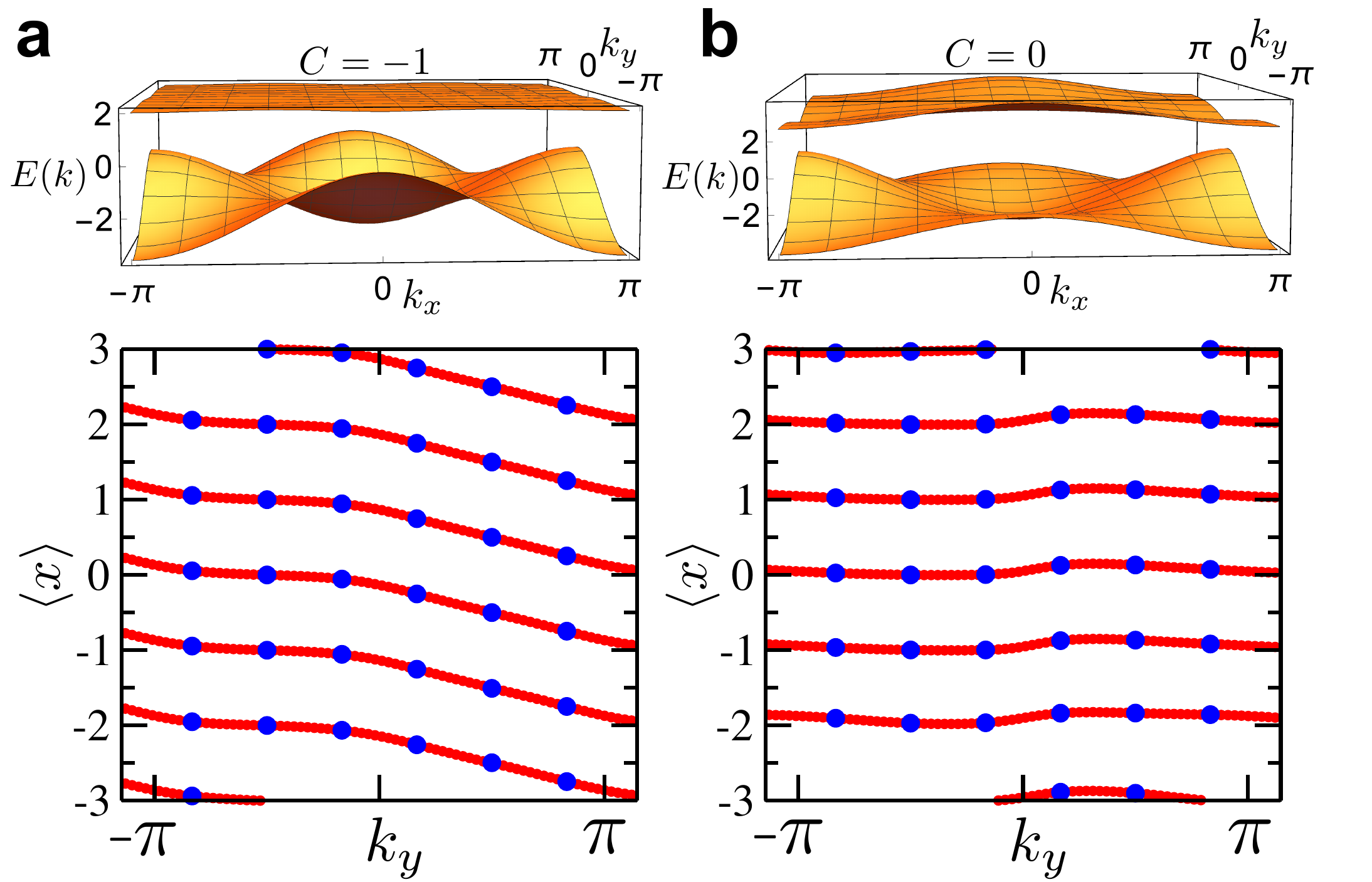}
\caption{
{\bf Topological winding of Wannier-Stark ladder eigenstates.}
Here, the effective electric field is applied along the $x$ direction. 
({\bf a}) Wave packet center position $\langle x \rangle$ of WSL eigenstates as a function of $k_y$ in the Chern flat band of the checkerboard lattice model, which is the upper energy band with the Chern number $C=-1$ as indicated in the top panel. 
Note that the topological winding with negative slope confirms $C=-1$.
({\bf b}) Similar plot in the topologically trivial situation, where all energy bands are topologically trivial with $C=0$.
As one can see, there is no winding.
Blue dots represent the wave packet center positions obtained at zero test flux, which are shifted along red lines with the insertion of the test flux. 
Here, $N_x=N_y=6$. 
}
\label{fig:winding}
\end{figure}

\noindent{\bf Topological basis.}
Qi has constructed the so-called maximally localized hybrid Wannier function (MLHWF) states~\cite{Qi11}, which can be one-to-one mapped to LL eigenstates in the Landau gauge. 
Actually, MLHWF states can be understood as Wannier-Stark ladder (WSL) eigenstates~\cite{Mendez88,Voisin88,Wilkinson96,Dahan96,Wacker02,Gluck02,Kruchinin18} formed in Chern flat bands under an effective electric field.
Specifically, WSL eigenstates are the energy eigenstates of the WSL Hamiltonian, i.e., ${\cal H}_{\rm WSL} \phi^{\rm WSL}={\cal E}_{\rm WSL} \phi^{\rm WSL}$ with
\begin{align}
{\cal H}_{\rm WSL}= \epsilon_{\bf k} +e{\bf E}\cdot(i\nabla_{\bf k}+{\cal A}_{\bf k}),
\label{eq:H_WSL}
\end{align}
where $\epsilon_{\bf k}$ is the energy dispersion of the given Chern flat band, ${\bf E}$ is the effective electric field tuning the wave packet width of WSL eigenstates, and ${\cal A}_{\bf k}= \langle u_{\bf k} | i \nabla_{\bf k} | u_{\bf k} \rangle$ is the Berry connection of Bloch states~\cite{Lee15}.
The energy eigenvalue of WSL eigenstates is given by
\begin{align}
{\cal E}_{\rm WSL}= \bar{\epsilon}(k_\perp) +\Omega \left( n +\gamma^{\rm Zak}(k_\perp)/2\pi \right),
\end{align}
where $k_\perp$ is the momentum perpendicular to ${\bf E}$, and $\Omega=eE a_\parallel$ is the Bloch frequency with $E$ and $a_\parallel$ being the strength of ${\bf E}$ and the lattice constant parallel to {\bf E}, respectively.
Similarly, $k_\parallel$ is the momentum parallel to ${\bf E}$, and $a_\perp$ is the lattice constant perpendicular to ${\bf E}$.
Also, $n$ is the WSL index, $\bar{\epsilon}(k_\perp)={a_\parallel \over 2\pi} \oint dk_\parallel  \epsilon_{\bf k}$ is the $k_\parallel$-averaged band energy, and 
$\gamma^{\rm Zak}(k_\perp) = \oint d {\bf k}_\parallel \cdot {\cal A}_{\bf k}$ is the celebrated Zak phase~\cite{Zak89}. 
Note that $\phi^{\rm WSL}$ can be expressed as $\phi^{\rm WSL}_{n,k_\perp}(k_\parallel)=\langle k_\parallel | \phi^{\rm WSL}_{n,k_\perp} \rangle$ with $| \phi^{\rm WSL}_{n,k_\perp} \rangle$ denoting the ket state:
\begin{align}
\phi^{\rm WSL}_{n,k_\perp}(k_\parallel)= e^{-\frac{i}{\Omega} \int^{k_\parallel}_0 dk_\parallel^\prime \left[ {\cal E}_{\rm WSL} -\epsilon_{{\bf k}^\prime} -\Omega \hat{e}_\parallel \cdot {\cal A}_{{\bf k}^\prime} \right]},
\end{align}
where ${\bf k}^\prime=k_\parallel^\prime \hat{e}_\parallel +k_\perp \hat{e}_\perp$ with $\hat{e}_\parallel$ and $\hat{e}_\perp$ being the unit vectors parallel and perpendicular to ${\bf E}$, respectively.
The MLHWF states can be obtained as the Fourier transform of $\phi^{\rm WSL}_{n,k_\perp}(k_\parallel)$ with respect to $k_\parallel$ in the limit of infinitely large $\Omega$, where the energy dispersion can be completely ignored. 
Generally, $\Omega$ can be treated as a variational parameter to tune the wave packet width of WSL eigenstates.

The wave packet center position of WSL eigenstates can be computed by taking the expectation value of the {\it gauge-invariant} position operator,  $a_\parallel(n+\gamma^{\rm Zak}(k_\perp)/2\pi)$, which is nothing but the polarization of WSL eigenstates~\cite{Vanderbilt93}. 
The Zak phase exhibits topological winding in the presence of non-trivial Chern numbers~\cite{Lee15}, meaning that the wave packet center position of WSL eigenstates is continuously displaced along $\hat{e}_\parallel$ as a function of $k_\perp$, scanning the entire system. 
See Fig.~\ref{fig:winding} for the illustration.
It is worthwhile to mention that the gauge-invariant position operator generates essentially the identical results to those obtained by using Resta's formula~\cite{Resta98}. 
This enables us to establish the one-to-one mapping between LL and WSL eigenstates via $k_{\rm LL} l_B/2\pi \leftrightarrow n +k_\perp a_\perp/2\pi$ with $k_{\rm LL}$ and $l_B$  being the LL momentum and the magnetic length, respectively.

Actually, in finite-size systems, the one-to-one mapping requires a proper implementation of the periodic (or twisted) boundary condition for both LL and WSL eigenstates.
For LL eigenstates, the periodic boundary condition can be implemented by using elliptic theta functions~\cite{Haldane85}.
For WSL eigenstates, we periodize them by discretizing $k_x$ via $k_x=2\pi p_x/N_x a_x$ with $p_x=1, \cdots, N_x$.
Concretely, ${\cal H}_{\rm WSL}$ is discretized by setting $[\epsilon_{\bf k}]_{{\bf k}_1{\bf k}_2}=\epsilon_{{\bf k}_1} \delta_{{\bf k}_1{\bf k}_2}$ and $[{\cal A}_{\bf k}]_{{\bf k}_1{\bf k}_2}={\cal A}_{{\bf k}_1} \delta_{{\bf k}_1{\bf k}_2}$, where $\delta_{{\bf k}_1{\bf k}_2}$ is the usual Kronecker delta.
Meanwhile, the momentum differentiation, or canonical position operator is replaced by $[\hat{\cal R}]_{{\bf k}_1{\bf k}_2} \equiv \langle{\bf k}_1|i\nabla_{\bf k}|{\bf k}_2\rangle= i\nabla_{{\bf k}_1} \delta_{\rm D}({\bf k}_1-{\bf k}_2)$, where $\delta_{\rm D}({\bf k}_1-{\bf k}_2)$ is the discretized delta function defined as the Dirichlet kernel~\cite{Kim16}.
Periodized WSL eigenstates can be obtained by diagonalizing the so-discretized WSL Hamiltonian.

Finally, we diagonalize ${\cal H}_{\rm CFB}$ by using periodized WSL eigenstates as basis states instead of usual Bloch states.
In this choice of the topological basis, the exact Coulomb ground state forming the triple-degenerate GSM, $|\Psi^{\rm Coul}_{m=1,2,3}\rangle$, can be directly compared with the Tao-Thouless state. 

\begin{figure}
\includegraphics[width=\columnwidth,angle=0]{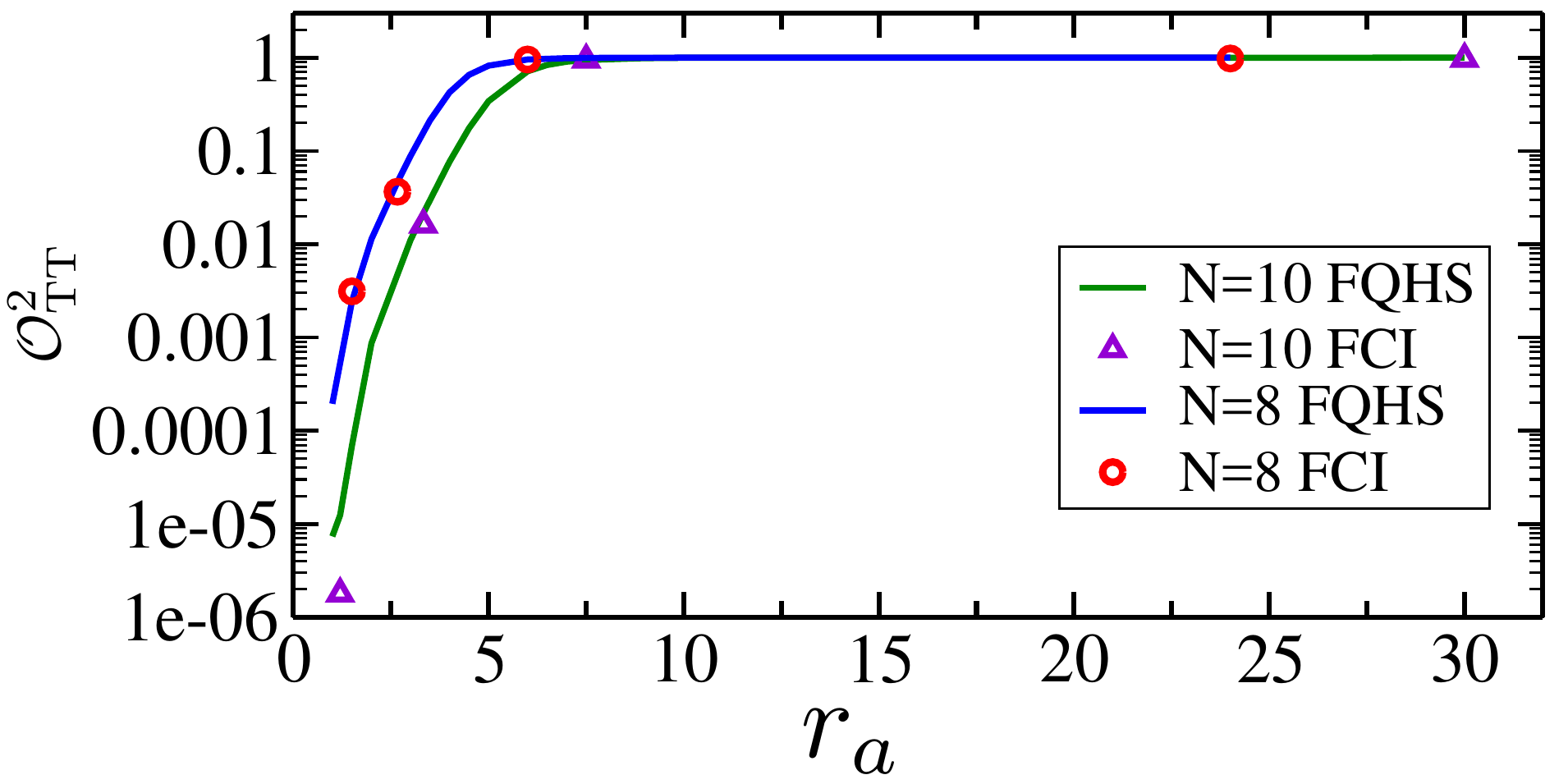}
\caption{
{\bf Overlap between fractional Chern insulators and the Tao-Thouless state.}
Open symbols denote the degeneracy-averaged square of overlap between the exact Coulomb ground and Tao-Thouless states, ${\cal O}^2_{\rm TT}$, in the 1/3-filled Chern flat band of the checkerboard lattice model as a function of $r_a=N_x/N_y$.
Continuous lines denote the similar overlap in the 1/3-filled LL. 
}
\label{fig:Tao-Thouless}
\end{figure}

\noindent{\bf Evolution to the Tao-Thouless state.}
The Tao-Thouless state can be written in the 1/3-filled LL as the following set of three root partition states represented in terms of LL eigenstates, $|100100\cdots\rangle, |010010\cdots\rangle, |001001\cdots\rangle$.
Based on the one-to-one mapping between LL and WSL eigenstates, the Tao-Thouless state can be written in the 1/3-filled Chern flat band as the similar set of root partition states now represented in terms of WSL eigenstates, $|\Psi^{\rm TT}_{n=1,2,3}\rangle$.
While the wave packet width of WSL eigenstates can be generally varied by tuning ${\bf E}$, here, we take the limit of large ${\bf E}$ so that WSL eigenstates become maximally localized, reducing to MLHWF states.
Later, we investigate what happens at finite ${\bf E}$.

Our goal is to check if $|\Psi^{\rm Coul}_{m=1,2,3}\rangle$ evolves to $|\Psi^{\rm TT}_{n=1,2,3}\rangle$ in the thin torus limit.
To this end, we compute the {\it degeneracy-averaged} square of overlap~\cite{Wu12_Gauge-fixed} between the two states,
${\cal O}^2_{\rm TT}=\frac{1}{3}\sum_{n,m} |\langle \Psi^{\rm TT}_n | \Psi^{\rm Coul}_m \rangle|^2$, as a function of $r_a$.
Figure~\ref{fig:Tao-Thouless} shows ${\cal O}^2_{\rm TT}$ in the 1/3-filled Chern flat band of the checkerboard lattice model as a function of $r_a$. 
Note that ${\cal O}^2_{\rm TT}$ follows almost exactly the behavior of the similar overlap in the 1/3-filled LL, becoming essentially unity at sufficiently large $r_a$, which confirms that the exact Coulomb ground state indeed evolves to the Tao-Thouless state.

\noindent{\bf Incorporating quantum fluctuations.} 
As one can see from Fig.~\ref{fig:Tao-Thouless}, while having the unity overlap in the thin torus limit, the exact Coulomb ground state has very low overlaps with the Tao-Thouless state at $r_a \simeq 1$.
For a better description of the 1/3-filling FCI in the 2D bulk, appropriate quantum fluctuations should be incorporated into the Tao-Thouless state. 
This essentially amounts to constructing the Chern-Laughlin state.

To this end, it is important to enforce the correct phase coherence among different WSL eigenstates so that they can form a coherently {\it gauge-fixed} basis~\cite{Wu12_Gauge-fixed}. 
To do so, here, we utilize the peculiar property of WSL eigenstates that they continuously evolve to each other via the topological winding as a function of $k_\perp$ just like LL eigenstates as a function of $k_{\rm LL}$. 
Utilizing this property, one can construct coherently gauge-fixed WSL eigenstates,
$| \phi^{\rm WSL}_{n,k_\perp} \rangle = e^{i\Theta_{n,k_\perp}} |\overline{\phi^{\rm WSL}_{n,k_\perp}} \rangle$ ,
where $|\overline{\phi^{\rm WSL}_{n,k_\perp}} \rangle$ is normalized under the particular gauge-fixing condition that
$\overline{\phi^{\rm WSL}_{n,k_\perp}}(k_\parallel)=\langle k_\parallel |\overline{\phi^{\rm WSL}_{n,k_\perp}} \rangle$ is real at a certain value of $k_\parallel$, which should be carefully chosen to avoid any singularities of ${\cal A}_{\bf k}$ along the line of $(k_\parallel, k_\perp)$ with $k_\perp \in [0,2\pi]$. 
The phase $\Theta_{n,k_\perp}$ is defined as $\Theta_{n,k_\perp} = \int^{2\pi n+k_\perp}_{0} d \kappa {\cal A}^{\rm WSL}(\kappa)$, where ${\cal A}^{\rm WSL}(\kappa)=\langle \overline{\phi^{\rm WSL}_{n,k_\perp^\prime}}| i\partial_{k_\perp^\prime} |\overline{\phi^{\rm WSL}_{n,k_\perp^\prime}} \rangle \Big|_{\kappa=2\pi n+k_\perp^\prime}$ is the Berry connection of WSL eigenstates.

The Chern-Laughlin state can be constructed by importing the amplitude of the Laughlin state in each LL eigenstate basis 
and attaching it to the corresponding WSL eigenstate basis connected via the one-to-one mapping. 
Denoting the triple-degenerate GSM of the Chern-Laughlin state as $| \Psi^{\rm CL}_{n=1,2,3} \rangle$, the overlap between the exact Coulomb ground and Chern-Laughlin states can be computed via ${\cal O}^2_{\rm CL}=\frac{1}{3}\sum_{n,m} |\langle \Psi^{\rm CL}_n | \Psi^{\rm Coul}_m \rangle|^2$.

\begin{figure}
\includegraphics[width=\columnwidth,angle=0]{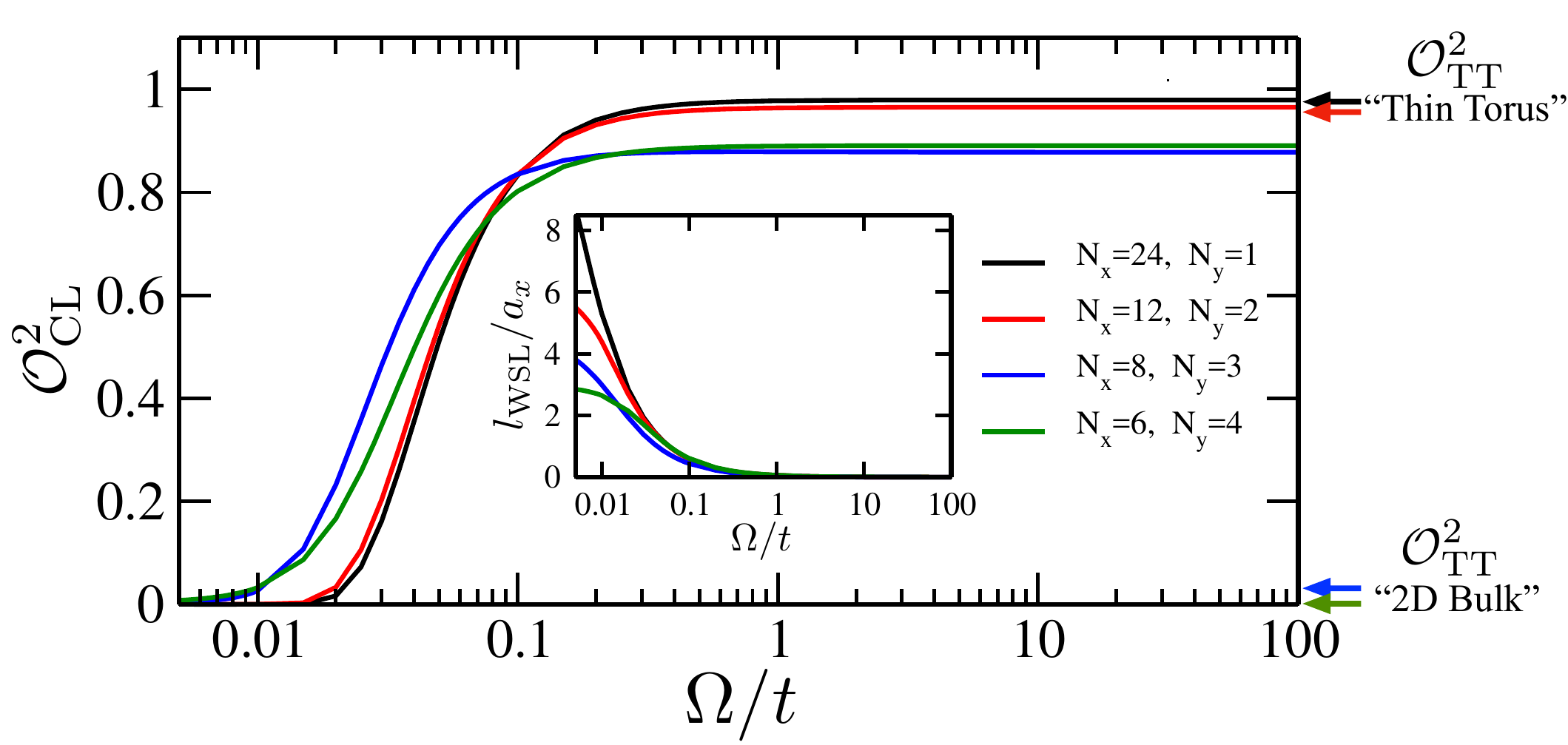}
\caption{
{\bf Overlap between fractional Chern insulators and the Chern-Laughlin state.}
Here, ${\cal O}^2_{\rm CL}$ denotes the degeneracy-averaged square of overlap between the exact Coulomb ground and Chern-Laughlin states in the 1/3-filled Chern flat band of the checkerboard lattice model as a function of the effective electric field strength, or the Bloch frequency $\Omega/t$.
The inset shows the wave packet width of WSL eigenstates, $l_{\rm WSL}/a_x$, as a function of $\Omega$, which can be computed via the standard deviation of the gauge-invariant position operator averaged over different WSL eigenstates. 
Color-coded arrows on the right side indicate ${\cal O}^2_{\rm TT}$ of the corresponding systems with different aspect ratios, which can be grouped as ``Thin Torus'' and ``2D Bulk.''
}
\label{fig:Chern-Laughlin}
\end{figure}

Figure~\ref{fig:Chern-Laughlin} shows ${\cal O}^2_{\rm CL}$ as a function of the effective electric field strength for different aspect ratios.
Three features are worth noticing. 
First, ${\cal O}^2_{\rm CL}$ is always maximized in the limit of large electric field strengths. 
This means that the exact Coulomb ground state is best described by the Chern-Laughlin state with maximally-localized WSL eigenstates. 
Second, quantum fluctuations take up a dominant portion of the exact Coulomb ground state in the 2D bulk as shown by the huge increase from ${\cal O}^2_{\rm TT}$ to ${\cal O}^2_{\rm CL}$ in ``2D Bulk'' systems.  
The existence of such large quantum fluctuations indicates that the FCI is not a simple charge density wave state.  
Third, the maximum of ${\cal O}^2_{\rm CL}$ is quite high not only in the thin torus limit, but also in the 2D bulk, proving that the exact Coulomb ground state is well described by the Chern-Laughlin state regardless of the aspect ratio.

Finally, we confirm that the Chern-Laughlin state provides an excellent description of the 1/3-filling FCI even for the nearest-neighbor interaction with basically the same level of the overlap for ``2D Bulk'' systems in Fig.~\ref{fig:Chern-Laughlin}, being consistent with previous results~\cite{Wu12_Gauge-fixed}.

\noindent{\bf Discussion}

In this work, we show that FCIs can be adiabatically connected to the Tao-Thouless state via the piecewise hybrid adiabatic path of first transforming the electron-electron interaction from the nearest-neighbor to Coulomb interaction, and then taking the thin torus limit. 
Being the lattice analogue of the Laughlin state, the Chern-Laughlin state can be constructed by incorporating appropriate quantum fluctuations into the Tao-Thouless state.

Our method can be extended to possible FCIs at other general fillings of the Jain sequence, 
where the mathematical forms of root partition states are explicitly known~\cite{Bergholtz08}.
The one-to-one mapping between LL and WSL eigenstates can be used to generate the lattice analogues of general CF states with the Chern CF sea being an exciting possibility.

\begin{acknowledgments}

The authors are grateful to Jainendra K. Jain for insightful discussions.
Also, the authors thank Center for Advanced Computation (CAC) at Korea Institute for Advanced Study (KIAS) for providing computing resources for this work.
This work is partially supported by the KIAS Individual Grants, PG034303 (SM) and PG032303 (KP).

\end{acknowledgments}



\end{document}


\title{Adiabatic Path from Fractional Chern Insulators to the Tao-Thouless State: Supplemental Material}

\author{Sutirtha Mukherjee}
\author{Kwon Park}
\affiliation{School of Physics, Korea Institute for Advanced Study, Seoul 02455, Korea}

\date{\today}

\maketitle

\noindent{\bf Tight-binding Models}

We focus on two tight-binding models with one in the checkerboard lattice~\cite{Sun11} and the other in the kagome lattice~\cite{Tang11}.

First, the Hamiltonian for the checkerboard lattice model is written as follows:
\begin{align}
H_{\rm checker} = \sum_{\bf k} 
( c_{a {\bf k}}^\dagger \;\; c_{b {\bf k}}^\dagger ) 
H_{\bf k} 
\left( 
\begin{array}{c}
c_{a {\bf k}} \\
c_{b {\bf k}} 
\end{array}
\right),
\label{H_checker1}
\end{align}
where
\begin{align}
H_{\bf k} &= -\left[ (t_1^\prime+t_2^\prime)(\cos{k_x} + \cos{k_y})+4t^{\prime\prime} \cos{k_x} \cos{k_y} \right] I 
\nonumber\\
&- 4t \cos{\phi} \cos{k_x \over 2} \cos{k_y \over 2} \sigma_x 
- 4t \sin{\phi} \sin{k_x \over 2} \sin{k_y \over 2} \sigma_y 
\nonumber\\
&- \left[ (t_1^\prime-t_2^\prime)(\cos{k_x} - \cos{k_y})+M \right] \sigma_z,
\label{eq:H_checker2}
\end{align} 
where $c_{a {\bf k}}$ and $c_{b {\bf k}}$ are the electron annihilation operators at momentum ${\bf k}$ for sub-lattices $a$ and $b$, respectively, $I$ is the identity matrix, and $\sigma_{x,y,z}$ are the Pauli matrices. 
As written above, $H_{\rm checker}$ is not of the proper Bloch form, which needs to be addressed in order to obtain the correct Chern number. 
$H_{\rm checker}$ can be brought to the proper Bloch form via the gauge transformation of $c_{b {\bf k}} \to e^{-i(k_x-k_y)/2} c_{b {\bf k}}$~\cite{Regnault11}.
Finally, the value of $M$ controls the phase transition between topologically trivial and non-trivial situations.

For the set of parameters with $t=1$, $t_1^\prime=-t_2^\prime={1/(2 + \sqrt{2})}$, $t^{\prime\prime} = 1/(2 + 2 \sqrt{2})$, $\phi = \pi/4$, and $M=0$, $H_{\rm checker}$ generates a gapped energy spectrum containing two energy bands with Chern numbers $C= \pm 1$. 
The upper energy band with $C= - 1$ has a bandwidth of nearly 1/30 of the band gap and thus can serve as a Chern flat band suitable for the realization of fractional Chern insulators (FCIs).

Second, the Hamiltonian for the kagome lattice model is written as follows:
\begin{align}
H_{\rm kagome} = \sum_{\bf k} 
( c_{a {\bf k}}^\dagger \;\; c_{b {\bf k}}^\dagger \;\;  c_{c {\bf k}}^\dagger ) 
H_{\bf k} 
\left( 
\begin{array}{c}
c_{a {\bf k}} \\
c_{b {\bf k}} \\
c_{c {\bf k}}
\end{array}
\right),
\label{eq:H_kagome1}
\end{align}
where
\begin{align}
H_{\bf k} = 
\left( 
\begin{array}{ccc}
0 & h_{12}({\bf k}) & h_{13}({\bf k}) \\
h_{12}^*({\bf k}) & 0 & h_{23}({\bf k}) \\
h_{13}^*({\bf k}) & h_{23}^*({\bf k}) & 0
\end{array}
\right)
\label{eq:H_kagome2}
\end{align}
with nonzero matrix elements of $H_{\bf k}$ given by
\begin{align}
h_{12}({\bf k}) &=  -2(t_1-i\lambda_1)\cos{k_1 \over 2} -2 (t_2+i\lambda_2)\cos{\left({k_2\over 2} + {k_3\over 2}\right)}  ,
\nonumber\\
h_{13}({\bf k}) &= -2(t_1+i\lambda_1)\cos{k_2 \over 2} -2(t_2-i\lambda_2)\cos{\left({k_3 \over 2}-{k_1 \over 2}\right)} ,
\nonumber\\
h_{23}({\bf k}) &=  -2(t_1-i\lambda_1)\cos{k_3 \over 2} -2 (t_2+i\lambda_2)\cos{\left({k_1 \over 2}+{k_2 \over 2}\right)} ,
\label{eq:H_kagome3}
\end{align} 
where $k_1 = k_x$, $k_2 = (k_x+\sqrt{3}k_y)/2$, and $k_3 = k_2-k_1$. 
Similar to $H_{\rm checker}$, $H_{\rm kagome}$ can be brought to the proper Bloch form via the gauge transformation of $c_{b {\bf k}} \to e^{-i k_1/2}c_{b {\bf k}}$ and  $c_{c {\bf k}} \to e^{-i k_2/2}c_{c {\bf k}}$.

For the set of parameters with $t_1=1$, $t_2=-0.3$, $\lambda_1 = 0.28$, and $\lambda_2 = 0.2$,  $H_{\rm kagome}$ generates a gapped energy spectrum containing three bands with Chern numbers $C=1,0,-1$. 
The bottom energy band with $C=1$ has a bandwidth of nearly 1/52 of the band gap against the middle energy band, serving as a Chern flat band suitable for the realization of FCIs.

\noindent{\bf Matrix elements of the electron-electron interaction}

We begin by writing the electron-electron interaction operator in the real space as follows:
\begin{align}
\hat{U}  =   {1 \over 2} \sum_{a,b} \sum^\prime_{i ,j}  U({\bf r}^a_i-{\bf r}^b_j) {\cal P}_{\rm CFB} |{\bf r}^a_i, {\bf r}^b_j \rangle \langle {\bf r}^a_i, {\bf r}^b_j | {\cal P}_{\rm CFB},
\label{eq:U_real}
\end{align}
where  $a, b$ and $i, j$ denote sublattice and unit cell indices, respectively. 
The prime indicates that the summation is restricted so that $i \neq j$ if $a = b$.
The ket state $| {\bf r}^a_i \rangle$ represents the position eigenstate at ${\bf r}^a_i$.

The projection operator to a given Chern flat band, ${\cal P}_{\rm CFB}$, modifies the position eigenstates as follows:  
\begin{align}
{\cal P}_{\rm CFB} | {\bf r}^a_i \rangle = \sum_{\bf k} 
u^{*}_a({\bf k})
e^{i \theta^a_{\bf k}} 
e^{i {\bf k} \cdot {\bf r}_i^a} 
| {\bf k} \rangle , 
\label{eq:r_projection}
\end{align}
where $u_a({\bf k})$ is the $a$-sublattice amplitude of the Bloch eigenstate with momentum ${\bf k}$ projected to the Chern flat band. 
The phase factor $e^{i \theta^a_{\bf k}}$ is due to the gauge transformation required to bring the Hamiltonian to the proper Bloch form.

By using the above momentum space representation of projected position eigenstates, we can rewrite the electron-electron interaction operator in the momentum space as follows:
\begin{align}
\hat{U}  =   \sum_{{\bf k}_1,{\bf k}_2,{\bf k}_3,{\bf k}_4}  \langle{\bf k}_1, {\bf k}_2| \hat{U} |{\bf k}_3, {\bf k}_4\rangle 
c^\dagger_{{\bf k}_1} c^\dagger_{{\bf k}_2} c_{{\bf k}_3} c_{{\bf k}_4} ,
\label{eq:U_momentum}
\end{align}
where $\langle {\bf k}_1, {\bf k}_2 | \hat{U} | {\bf k}_3, {\bf k}_4 \rangle$ is the matrix element of the electron-electron interaction in terms of momentum eigenstates:
\begin{widetext}
\begin{align}
\langle {\bf k}_1, {\bf k}_2 | \hat{U} | {\bf k}_3, {\bf k}_4 \rangle = 
\sum_{a, b} \sum_{\delta{\bf r}} 
\delta_{{\bf k}_1+{\bf k}_2, {\bf k}_3+{\bf k}_4} 
U( \delta{\bf r} )  
e^{i({\bf k}_2-{\bf k}_4) \cdot \delta{\bf r}} 
e^{i (\theta^a_{{\bf k}_1}+\theta^b_{{\bf k}_2}-\theta^a_{{\bf k}_3}-\theta^b_{{\bf k}_4})} 
u^*_a ({\bf k}_1) u^*_b ({\bf k}_2) u_a ({\bf k}_3) u_b ({\bf k}_4) ,
\label{eq:U_matrix_element_momentum}
\end{align}
\end{widetext}
where $\delta{\bf r}={\bf r}^a_i-{\bf r}^b_j$ is the relative position vector.

Meanwhile, we can still rewrite the electron-electron interaction operator via the Wannier-Stark ladder (WSL) eigenstate representation as follows:
\begin{align}
\hat{U}  =   \sum_{\kappa_i,\kappa_j,\kappa_k,\kappa_l}  \langle \kappa_i, \kappa_j | \hat{U} | \kappa_k, \kappa_l \rangle 
c^\dagger_{\kappa_i} c^\dagger_{\kappa_j} c_{\kappa_k} c_{\kappa_l} ,
\label{eq:U_WSL}
\end{align}
where $\langle \kappa_i, \kappa_j | \hat{U} | \kappa_k, \kappa_l \rangle$ is the matrix element of the electron-electron interaction in terms of WSL eigenstates:
\begin{widetext}
\begin{align}
\langle \kappa_i, \kappa_j | \hat{U} | \kappa_k, \kappa_l \rangle &=
\sum_{{\bf k}_1, {\bf k}_2, {\bf k}_3, {\bf k}_4} 
\langle {\bf k}_1, {\bf k}_2 | \hat{U} | {\bf k}_3, {\bf k}_4 \rangle
\phi^*_{\kappa_i}({\bf k}_1) \phi^*_{\kappa_j}({\bf k}_2)  \phi_{\kappa_k} ({\bf k}_3) \phi_{\kappa_l}({\bf k}_4),
\label{eq:U_matrix_element_WSL}
\end{align}
\end{widetext}
where $\phi_{\kappa}({\bf k})$ is related with WSL eigenstates via $\phi_{\kappa}({\bf k})= \phi^{\rm WSL}_{n,k_\perp}(k_\parallel)$ with $\kappa=2\pi n+k_\perp$ and ${\bf k}=k_\parallel \hat{e}_\parallel+k_\perp \hat{e}_\perp$.
See the main text for the explicit definition of $\phi^{\rm WSL}_{n,k_\perp}(k_\parallel)$.

The physical form of the electron-electron interaction can be varied by changing $U(\delta {\bf r})$ in Eq.~\eqref{eq:U_matrix_element_momentum}. 
In the case of the nearest-neighbor interaction, $U(\delta {\bf r})$ is chosen to be $U^{\rm NN}(\delta {\bf r})$, which is nonzero only if $\delta {\bf r}$ is the relative position vector connecting between nearest-neighboring sites. 
In the case of the Coulomb interaction, $U(\delta {\bf r})$ is chosen to be $U^{\rm Coul}(\delta {\bf r})$, which is the usual $1/r$ potential except for an important modification due to the existence of image charges induced by the periodic boundary condition.

Specifically, following Bonsall and Maradudin~\cite{Bonsall77}, we can write $U^{\rm Coul}(\delta {\bf r})$ as follows:
\begin{widetext}
\begin{align}
U^{\rm Coul}(\delta{\bf r}) = 
\sum_{{\bf R}_l} {e^2 \over | \delta{\bf r}+{\bf R}_l |} 
={e^2 \over \sqrt{a_c}} \sum_{{\bf G} \neq 0}  e^{-i {\bf G} \cdot \delta{\bf r}} \varphi_{-{1 \over 2}}{\left( {|{\bf G}|^2 a_c \over 4\pi} \right)} 
+{e^2 \over \sqrt{a_c}} \sum_{{\bf R}_l} \varphi_{-{1\over 2}}{\left( {\pi \over a_c} |\delta{\bf r}+{\bf R}_l |^2  \right)} ,
\end{align} 
\end{widetext}
where ${\bf R}_l  = l_1 N_1 \hat{e}_1+l_2 N_2 \hat{e}_2$ with $\hat{e}_1$ and $\hat{e}_2$ being the primitive unit cell vectors of the underlying lattice, ${\bf G}$ is the corresponding reciprocal lattice vectors, $a_c$ is the area of the unit cell, and $\varphi_n(x)=\int_1^{\infty} dt~{t^n} e^{-xt}$.

\begin{figure*}
\includegraphics[width=1.8\columnwidth,angle=0]{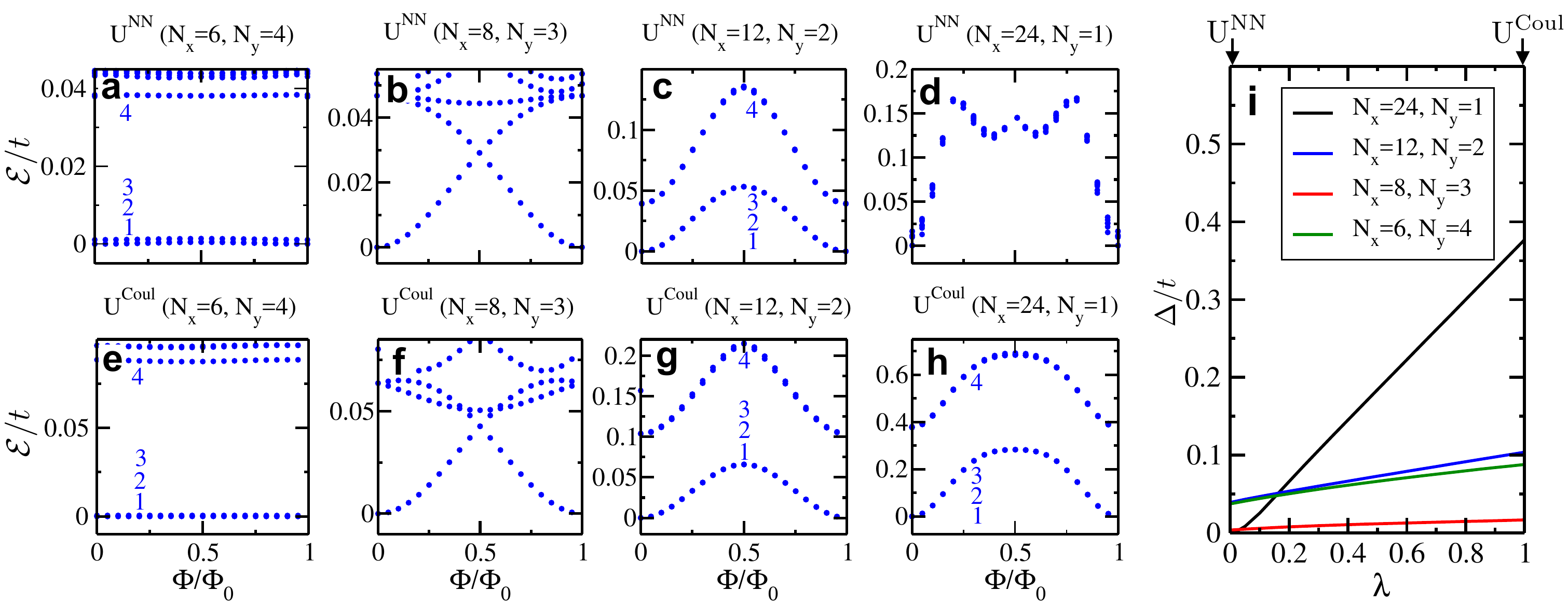}
\caption{ 
{\bf Comparison between the energy spectra of the nearest-neighbor and Coulomb interactions.}
({\bf a}\mbox{-}{\bf d}) Energy spectra of the nearest-neighbor interaction, $U^{\rm NN}$, in the Chern flat band of the kagome lattice model as a function of the test flux, $\Phi/\Phi_0$, for various different values of $(N_x,N_y)$.
$\Phi_0$ is the magnetic flux quantum.
({\bf e}\mbox{-}{\bf h}) Similar energy spectra of the Coulomb interaction, $U^{\rm Coul}$.
Here, the electron and site numbers are $N=8$ and $N_s=N_x N_y=24$, respectively.
Energies are given in units of the hopping amplitude $t_1$ and offset by the lowest energy in each panel.
The first four lowest energies are specified by numbers when three degenerate copies of the ground state have almost the same energy. 
({\bf i}) Energy gap $\Delta$ between the third and fourth lowest energies as a function of the mixing parameter $\lambda$ tuning the range of interaction via $U(\lambda)=(1-\lambda)U^{\rm NN}+\lambda U^{\rm Coul}$.
Here, $\Phi$ is set to be zero.
}
\label{fig:spectrum_kagome}
\end{figure*}

\noindent{\bf Results for the kagome lattice model}

In this section, we provide results for the kagome lattice model, which mirror those of the checkerboard lattice model.

\noindent{\bf Adiabatic path.}
Figure~\ref{fig:spectrum_kagome} shows the comparison between the energy spectra of the nearest-neighbor and Coulomb interactions in the kagome lattice as a function of $(N_x,N_y)$ for $N=8$ and $N_s=24$.
Specifically, it is shown in Fig.~\ref{fig:spectrum_kagome}~{\bf a}\mbox{-}{\bf d} that the triple-degenerate ground state manifold (GSM) collapses if one takes the thin torus limit with the nearest-neighbor interaction. 
On the other hand, it is shown in Fig.~\ref{fig:spectrum_kagome}~{\bf e}\mbox{-}{\bf h} that the triple-degenerate GSM remains intact in the case of the Coulomb interaction all the way through the most thin torus limit of the given finite-size system.

Similar to what is done for the checkerboard lattice model, we investigate the adiabatic continuity of the triple-degenerate GSM in the 2D bulk by transforming the electron-electron interaction from the nearest-neighbor to Coulomb interaction via $U(\lambda)=(1-\lambda)U^{\rm NN}+\lambda U^{\rm Coul}$.
Figure~\ref{fig:spectrum_kagome}~{\bf i} shows the energy gap between the third and fourth lowest energies as a function of $\lambda$, which remains intact in the 2D bulk, i.e., at $(N_x,N_y)=(6,4)$, while collapsing in the thin torus limit, i.e., at $(N_x,N_y)=(24,1)$.
Note that exactly the same behavior has been obtained in the checkerboard lattice model.

\begin{figure}
\includegraphics[width=\columnwidth,angle=0]{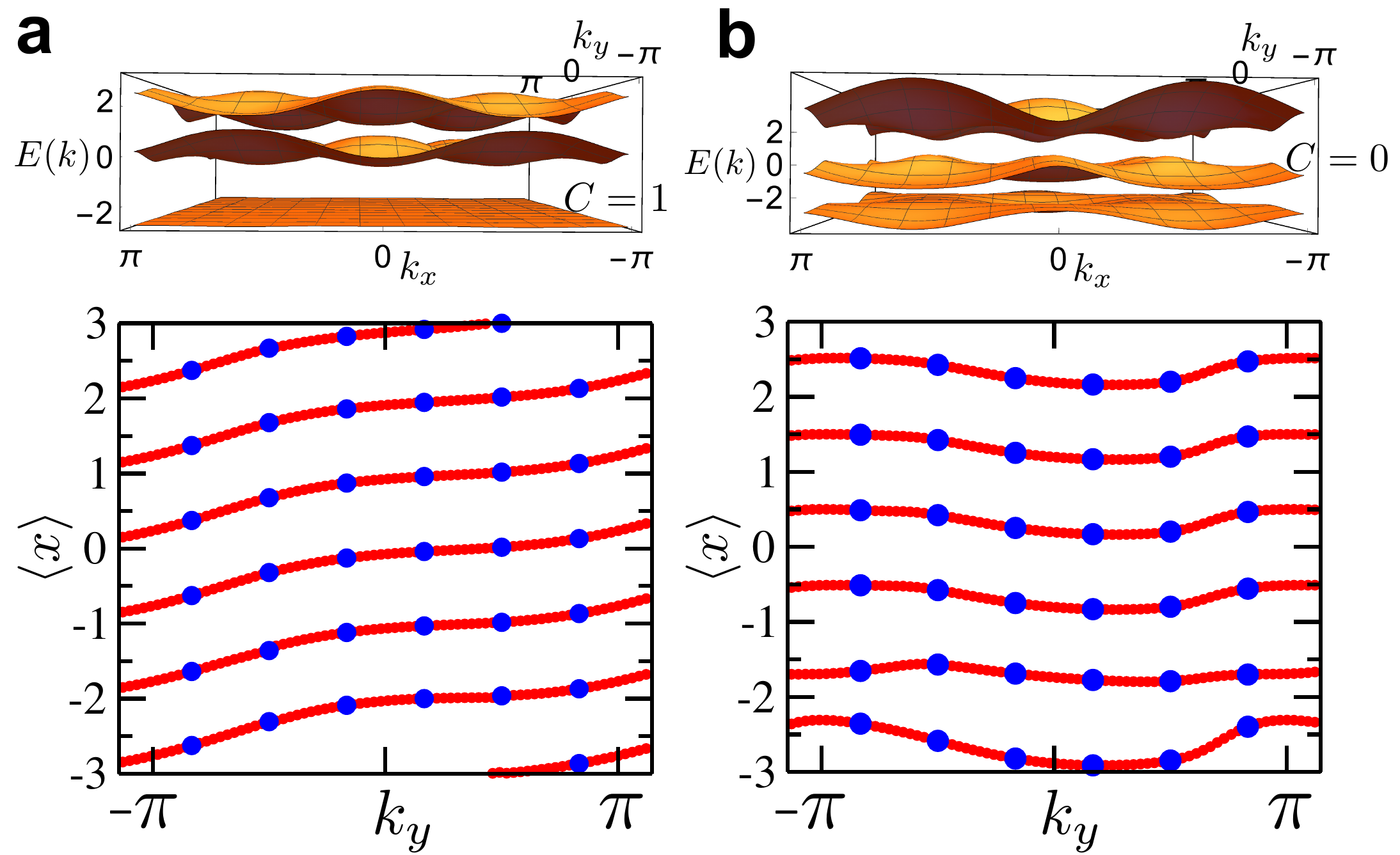}
\caption{
{\bf Topological winding of Wannier-Stark ladder eigenstates.}
Here, the effective electric field is applied along the $x$ direction. 
({\bf a}) Wave packet center position $\langle x \rangle$ of WSL eigenstates as a function of $k_y$ in the Chern flat band of the kagome lattice model, which is the bottom energy band with the Chern number $C=1$ as indicated in the top panel.
Note that the topological winding with positive slope confirms $C=1$.
({\bf b}) Similar plot in the topologically trivial band, which is the middle energy band with $C=0$ as indicated in the top panel.
As one can see, there is no winding.
Blue dots represent the wave packet center positions obtained at zero test flux, which are shifted along red lines with the insertion of the test flux. 
Here, $N_x=N_y=6$. 
}
\label{fig:winding_kagome}
\end{figure}

\noindent{\bf Topological basis.}
Figure~\ref{fig:winding_kagome} shows the topological winding of WSL eigenstates in the kagome lattice, which manifests itself via the wave packet center position of WSL eigenstates as a function of the conserved momentum.

Specifically, it is shown in Fig.~\ref{fig:winding_kagome}~{\bf a} that the wave packet center position of WSL eigenstates winds up as a function of the conserved momentum in the Chern flat band. 
On the other hand, it is shown in Fig.~\ref{fig:winding_kagome}~{\bf b} that there is no winding in the the topological trivial band.

\begin{figure}
\includegraphics[width=\columnwidth, angle=0]{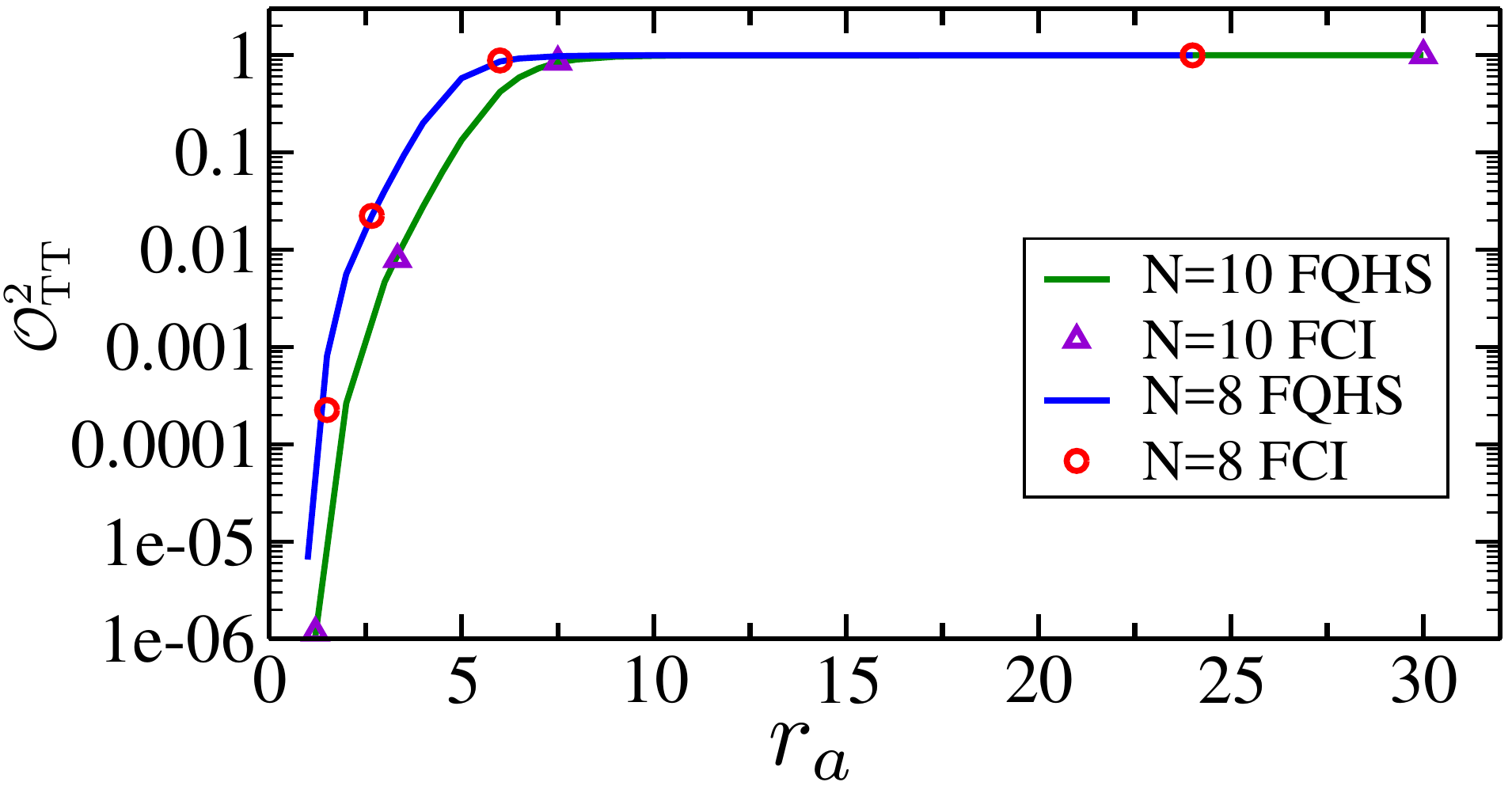}
\caption{
{\bf Overlap between fractional Chern insulators and the Tao-Thouless state.}
Open symbols denote the degeneracy-averaged square of overlap between the exact Coulomb ground and Tao-Thouless states, ${\cal O}^2_{\rm TT}$, in the 1/3-filled Chern flat band of the kagome lattice model as a function of $r_a$.
Continuous lines denote the similar overlap in the 1/3-filled LL.
Note that, here, LL eigenstates are constructed in the torus geometry with oblique unit cells, which match the shape of those in the kagome lattice. 
}
\label{fig:Tao-Thouless_kagome}
\end{figure}

\noindent{\bf Evolution to the Tao-Thouless state.}
Figure~\ref{fig:Tao-Thouless_kagome} shows the degeneracy-averaged square of overlap between the exact Coulomb ground and Tao-Thouless states, ${\cal O}^2_{\rm TT}$, in the 1/3-filled Chern flat band of the kagome lattice model as a function of $r_a$.

As one can see, the behavior of ${\cal O}^2_{\rm TT}$ in the kagome lattice model is almost exactly the same as that in the checkerboard counterpart. 
That is, first, ${\cal O}^2_{\rm TT}$ becomes essentially unity at sufficiently large $r_a$, confirming that the exact Coulomb ground state evolves to the Tao-Thouless state even in the kagome lattice model.
Second, ${\cal O}^2_{\rm TT}$ follows almost exactly the behavior of the similar overlap in the 1/3-filled LL, which is constructed in the torus geometry with oblique unit cells. 
Again, it is shown that the 1/3-filling FCI behaves just like the Laughlin state in the kagome lattice model.

\begin{figure}
\includegraphics[width=\columnwidth, angle=0]{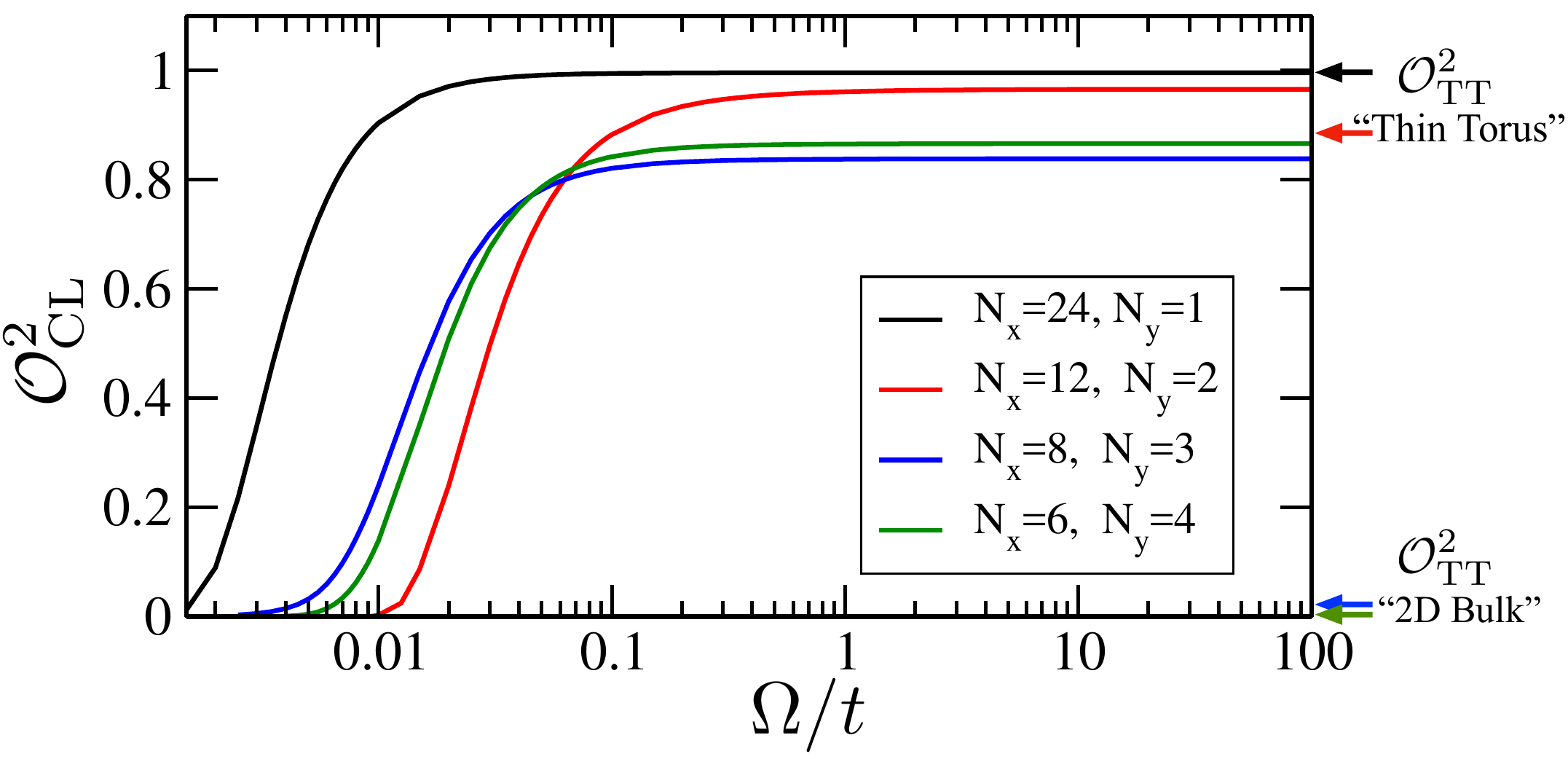}
\caption{
{\bf Overlap between fractional Chern insulators and the Chern-Laughlin state.}
Here, ${\cal O}^2_{\rm CL}$ denotes the degeneracy-averaged square of overlap between the exact Coulomb ground and Chern-Laughlin states in the 1/3-filled Chern flat band of the kagome lattice model as a function of the effective electric field strength, or the Bloch frequency $\Omega/t$.
Color-coded arrows on the right side indicate ${\cal O}^2_{\rm TT}$ of the corresponding systems with different aspect ratios, which can be grouped as ``Thin Torus'' and ``2D Bulk.''
}
\label{fig:Chern-Laughlin_kagome}
\end{figure}

\noindent{\bf Incorporating quantum fluctuations.}
Figure~\ref{fig:Chern-Laughlin_kagome} shows the degeneracy-averaged square of overlap between the exact Coulomb ground and Chern-Laughlin states, ${\cal O}^2_{\rm CL}$, in the 1/3-filled Chern flat band of the kagome lattice model as a function of $\Omega/t$ for different aspect ratios.

Here, essentially the same three features are obtained as those obtained in the checkerboard lattice model.
First, ${\cal O}^2_{\rm CL}$ is always maximized in the limit of $\Omega$, meaning that the exact Coulomb ground state is best described by the Chern-Laughlin state with maximally-localized WSL eigenstates even in the kagome lattice model. 
Second, quantum fluctuations take up a dominant portion of the exact Coulomb ground state in the 2D bulk.
Third, regardless of the aspect ratio, the exact Coulomb ground state is well described by the Chern-Laughlin state, which is constructed via incorporating appropriate quantum fluctuations into the Tao-Thouless state.

Finally, it is confirmed in the kagome lattice that, in the 2D bulk, the Chern-Laughlin state still provides an excellent description of the 1/3-filling FCI even for the nearest-neighbor interaction with basically the same level of the overlap for ``2D Bulk'' systems in Fig.~\ref{fig:Chern-Laughlin_kagome}.

We would like to mention that our overlap values are slightly different from those previously obtained in the same kagome lattice model, while basically the same in the checkerboard counterpart~\cite{Wu12_Gauge-fixed}.
This discrepancy is mainly due to the fact that, in the case of the kagome lattice model, we have used a different set of system parameters in the Hamiltonian from that in previous studies.
We have chosen our set of system parameters to make the energy dispersion of the Chern flat band as flat as possible rather than simply ignoring it.

